%% file: main.tex
\documentclass[sigconf]{acmart}
\AtBeginDocument{%
  }

\setcopyright{acmlicensed}
\copyrightyear{2018}
\acmYear{2018}
\acmDOI{XXXXXXX.XXXXXXX}
\acmConference[Conference acronym 'XX]{Make sure to enter the correct
  conference title from your rights confirmation email}{June 03--05,
  2018}{Woodstock, NY}
\acmISBN{978-1-4503-XXXX-X/2018/06}

\usepackage{placeins} 
\usepackage{makecell}
\usepackage{listings}
\usepackage{xcolor}

\begin{document}

\title{A Systematic Study of LLM-Based Architectures for Automated Patching}


\author{Qingxiao Xu, Ze Sheng, Zhicheng Chen, Jeff Huang}
\affiliation{%
  \institution{Texas A\&M University}
  \city{College Station}
  \country{USA}
}
\email{%
  {qingxiao,zesheng,chenzc2001,jeff}@tamu.edu
}


\begin{abstract}
Large language models (LLMs) have shown promise for automated patching of software security vulnerabilities, but their effectiveness depends strongly on how they are integrated into patching systems. While prior work explores prompting strategies and individual agent designs, the field lacks a systematic comparison of patching architectures. In this paper, we present a controlled evaluation of four LLM-based patching paradigms, including fixed workflow, single-agent system, multi-agent system, and general-purpose code agents, using a unified benchmark and evaluation framework. We analyze patch correctness, tool usage, token usage, and execution time across real-world vulnerability tasks.
\textit{\textbf{Surprisingly, general-purpose code agents achieve the strongest overall patching performance}}, benefiting from general-purpose tool interfaces that support effective adaptation across vulnerability types. Moreover, our results reveal clear architectural trade-offs: fixed workflows are efficient but brittle, single-agent systems balance flexibility and cost, and multi-agent designs improve generalization at the expense of substantially higher overhead and increased risk of reasoning drift on complex tasks. Overall, we show that architectural design and iteration depth, rather than model capability alone, dominate the reliability and cost of LLM-based automated patching.
\end{abstract}




\received{20 February 2007}
\received[revised]{12 March 2009}
\received[accepted]{5 June 2009}

\maketitle

\section{Introduction}
Modern software ecosystems amplify the impact of vulnerabilities: a single flaw such as Log4Shell (CVE-2021-44228) in Apache Log4j or the recent MongoBleed vulnerability (CVE-2025-14847) in MongoDB, can cascade through dependency chains and compromise large numbers of downstream systems~\cite{log4shell, mongobleed}. While vulnerability detection is increasingly automated~\cite{zhou2024large, guo2024outside, sheng2025llms}, {\em repair remains the dominant bottleneck}: patches must preserve program semantics, eliminate exploitability, and integrate cleanly into large, evolving codebases. Fully automating this process is difficult because it requires deep semantic reasoning, precise fault localization, and rigorous validation, making scalable and correct vulnerability remediation an open challenge.

Recent advances in large language models (LLMs) have begun to break this bottleneck. In DARPA's AI Cyber Challenge (AIxCC)~\cite{darpa_aicyberchallenge_2024}, finalist teams demonstrated fully automated Cyber Reasoning Systems (CRSs) for {\em real-world} open-source projects that not only detected 77\% (54/70) of injected vulnerabilities, but also correctly patched 61\% (43/70) of them. Subsequent systems such as DeepMind's CodeMender~\cite{deepmind_codemender2025} and OpenAI's Aardvark~\cite{openai_aardvark2025} further showed that LLM-driven pipelines (combined with fuzzing, symbolic execution, and formal verification) can generate human-quality, upstreamable patches. Together, these results mark a fundamental shift: {\em vulnerability repair is no longer purely manual}, but increasingly driven by autonomous, reasoning-centric AI systems.

Despite this progress, critical questions remain unanswered. Although all AIxCC finalist CRSs leverage LLMs to generate patches, there is little understanding of how different architectural integration strategies affect repair effectiveness, efficiency, and robustness. Existing work~\cite{zhang2024autocoderover, wang2024openhands, yang2024swe, lee2024unified, zhang2024poster} overwhelmingly treats system architecture as an implementation detail rather than a first-class design variable. Moreover, leading production systems (i.e., CodeMender and Aardvark) are proprietary, obscuring critical architectural decisions and making it difficult to assess whether their design choices generalize across vulnerability classes and codebases.

In this paper, we address this gap by conducting the first architecture-centric study of LLM-based vulnerability patching systems. We identify and analyze four distinct architectural paradigms: three domain-specific patching architectures including {\em fixed-workflow} pipelines, {\em single-agent} systems and {\em multi-agent} systems, and {\em general-purpose coding} agents, which collectively capture the core architectural patterns used in the patching components of the seven AIxCC finalist CRSs; some systems adopt a single pattern, while others combine multiple architectures. To enable controlled comparison, we reimplement the first three architectures within a unified patching framework~\footnote{Our code has been released as open source on \url{https://anonymous.4open.science/r/Patch-Architectures-07C6}.} and evaluate them on a collection of 19 delta-scan Java vulnerabilities from the AIxCC benchmark, including six large, complex, real-world Java projects (i.e., Apache Commons Compress, ZooKeeper,  Log4j, Tika, PDFBox, and POI). 
We evaluate the fourth paradigm using Claude Code~\cite{anthropic_claudecode_2025} as a representative general coding agent.

Our results reveal clear trade-offs across different architectures for automated vulnerability repair. Surprisingly, general coding agents achieve the strongest overall patching performance and robustness, particularly on large codebases, but at significantly higher token cost. Claude Code successfully repairs 16/19 vulnerabilities, while the best patch-specific agent repairs at most 13/19 (multi-agent on GPT-5). Patch-specific agent architectures are also substantially more token-efficient and faster, yet more prone to brittle search behavior and vulnerability types as reasoning scope expands. Notably, multi-agent designs do not consistently outperform well-designed single-agent systems, indicating that coordination overhead can offset the benefits of agent specialization in vulnerability repair. These findings show that repair effectiveness depends not only on model capability, but critically on how reasoning, control, and tool usage are architected.

In sum, our contributions are:
\begin{itemize}
  \item \textbf{Architecture-level taxonomy.} We identify and formalize four architectural paradigms for LLM-based vulnerability patching, and present their design and implementation in detail in Section~3. To the best of our knowledge, this is the first work to clearly describe, reimplement, and systematically evaluate these architectures under a unified experimental setting.
  \item \textbf{Unified evaluation on the AIxCC dataset.} We develop a unified evaluation framework and compare representative systems across all four architectures on the AIxCC benchmark. Our results show that general coding agents achieve the best overall performance, while other architectures remain important by successfully patching complementary cases.
  \item \textbf{System behavior and cost analysis.} We conduct an in-depth log-based analysis of system behavior, comparing tool usage, agent interactions, cost, and latency. We find that patch-specific agents are more token-efficient, while general coding agents show stronger robustness against infinite loops and long-running failures.
\end{itemize}

\section{Background and Problem Setting}

\subsection{Automated Vulnerability Repair}

Automated Program Repair (APR)~\cite{le2019automated} aims to synthesize source-level patches that fix general software defects without introducing regressions. In the context of security, APR focuses on \emph{vulnerability repair}: eliminating exploitable behaviors while preserving intended program semantics. Given the high cost and scalability limits of manual remediation, vulnerability repair has become an important problem in secure software maintenance.

A substantial body of prior work addresses APR using both program analysis-driven techniques and LLM–based approaches~\cite{huang2024evolving}. We review representative systems and techniques in Section~\ref{sec:relatedwork}.

In a typical vulnerability repair setting, the system is provided with (i) a target program, (ii) evidence of a vulnerability, such as a crashing input, exploit trigger, or static analysis report, and (iii) optional validation mechanisms, including tests or runtime checks. The objective is to generate a patch that removes the vulnerability, passes validation, and integrates cleanly into the existing codebase. 
Unlike functional bug repair, security patching must reason about exploitability, adversarial inputs, and subtle control- and data-flow properties. As a result, correctness, robustness, and resistance to incomplete fixes are central concerns.

\subsection{The AIxCC Patching Task}

DARPA has continuously organized competitions that emphasize end-to-end automation, with the goal of minimizing human intervention in the defense process. In the 2013–2016 Cyber Grand Challenge (CGC)~\cite{darpa_cgc}, representative systems such as Mayhem~\cite{6234425} and Mechanical Phish~\cite{stephens2016driller} operated purely at the binary level and were primarily driven by fuzzing and symbolic execution techniques. These systems established the paradigm of fully autonomous vulnerability discovery and patching, but were primarily built upon rigid, highly engineered analysis pipelines.

The DARPA AI Cyber Challenge (AIxCC)~\cite{darpa_aicyberchallenge_2024}, instead of binary-only analysis, operates at the source-code level and targets large, real-world software systems, providing a realistic and demanding benchmark for automated vulnerability repair. In this work, we focus specifically on the \emph{patching task} as instantiated in the AIxCC delta-scan Java vulnerability dataset, with the aim of understanding how LLMs can be effectively leveraged for reasoning, planning, and tool orchestration to build generalizable patching systems.

For each vulnerability, the patching system is provided with (i) the full source code of the target program, (ii) a complete \emph{proof of vulnerability (PoV)} packaged in a containerized environment, (iii) the fuzzer harness source code, and (iv) the functionality test scripts. PoV typically contains a crashing input that triggers the vulnerability, produced by a Jazzer-based fuzzer. All artifacts are provided within a Docker environment that enables deterministic reproduction of the failure.

In addition, for delta-scan tasks, the system is provided with the \emph{commit diff} that introduced the vulnerability. This diff captures the code changes responsible for the crashing behavior and serves as an explicit signal for fault localization, while still requiring the system to understand the full source code and reason about the root cause.

The goal of the patching system is to generate a \emph{source-level patch}, expressed in unified diff format, that eliminates the crashing behavior. A candidate patch is considered correct if it both (i) prevents the crash under the provided PoV and (ii) passes a project-specific functional test suite. This script is invoked by the patching system to test patch correctness, but provides no additional semantic guidance.

The delta-scan vulnerabilities evaluated in this work are intentionally non-trivial: they often span multiple files, require understanding of project-specific abstractions, and cannot be resolved through simple syntactic edits. This realistic difficulty, combined with correctness criteria based on security and functionality, makes the benchmark a rigorous evaluation setting for automated vulnerability repair.


\input{arch}

\section{Results}
We evaluate the different patching architectures using the official dataset from the DARPA AI Cyber Challenge (AIXCC)~\cite{darpa_aicyberchallenge_2024}.
Our experiments focus on the 19 Java projects evaluated under AIXCC’s delta-scan mode. In this setting, systems are given access to incremental code changes (deltas) between successive project versions, reflecting a realistic patching scenario in which vulnerabilities are introduced by recent commits. This mode emphasizes precise localization and minimal corrective edits, and is therefore well suited for comparing architectural differences in patch systems.


For each architectural design, we benchmark patching performance using two state-of-the-art large language models: GPT-5 and Claude Sonnet-4.5, \footnote{Both models were released before the public release of the AIXCC competition data, ensuring that their training data does not include any AIXCC competition artifacts.}. To evaluate general-purpose coding agents that are not tailored specifically for vulnerability repair, we additionally include Claude Code (also based on Claude Sonnet-4.5). This setup allows us to contrast domain-specific patching architectures with general coding agents under a unified and controlled evaluation framework. 


\subsection{Patch Correctness}
The results (Table~\ref{tab:results}) show a clear trend: the general code agent achieves the strongest performance, outperforming both agentic systems and workflow system. This suggests that domain-specific tool orchestration—where the model explicitly plans how to call customized tools—may impose rigidity that limits flexibility during patch generation. In contrast, Claude Code benefits from a richer, adaptive planning loop, dynamically updating its internal TODO list and interacting with the codebase in a more opportunistic, developer-like manner. These findings imply that highly general, interactive agents with fine-grained autonomy may ultimately surpass custom-designed APR workflows, even those carefully optimized for the patching domain.

For Claude Code, we observed a mismatch between self-reported success and actual test outcomes. Claude Code reported success for every challenge. Therefore, we independently executed the full test suites and identified three cases where Claude Code claimed success but the patched programs either failed functional tests or modified test files.

\begin{table}[h]
\centering
\begin{tabular}{lcc}
\hline
\textbf{System} & \textbf{GPT-5} & \textbf{Sonnet-4.5} \\
\hline
Single Agent     & 12/19 & 13/19 \\
Multi-Agent      & 13/19 & 8/19  \\
Fuzzing-Brain    & 12/19 & 10/19 \\
ClaudeCode       & --    & 16/19 \\
\hline
\end{tabular}
\caption{Comparison of patch success rates across architectures and models.}
\label{tab:results}
\end{table}

\subsubsection{\bf A Failure Case of Claude Code}
Although Claude Code achieved the highest overall patch success rate, it failed on two challenges in the Tika and ZooKeeper projects. In contrast, these vulnerabilities were successfully patched under certain agentic-system and workflow-based configurations. This indicates that systems which perform worse than Claude Code can still be effective in specific cases, acting as complementary approaches. These results highlight inherent limitations of general coding agents and motivate the continued exploration of specialized patching architectures.

One failure case in Claude Code is a ZooKeeper challenge, an infinite-loop denial-of-service vulnerability introduced by the newly added IPv6 validation logic in \texttt{MessageTracker}. In this case, the vulnerability manifests as an unbounded loop when processing malformed server address strings, allowing an attacker to trigger excessive iteration and denial of service.

Claude Code failed to remediate this vulnerability, whereas a multi-agent GPT-5 configuration produced a valid patch, as shown in Listings~\ref{lst:zk_claude_patch} and~\ref{lst:zk_gpt5_patch}. This behavior highlights a limitation of general coding agents: the agent prematurely concluded that its initial modification was sufficient and terminated before exhaustively validating the patch against all PoVs.

\begin{lstlisting}[language=, caption={Claude Code patch}, label={lst:zk_claude_patch}]
diff --git a/zookeeper-server/src/main/java/org/apache/zookeeper/server/util/MessageTracker.java b/zookeeper-server/src/main/java/org/apache/zookeeper/server/util/MessageTracker.java
--- a/zookeeper-server/src/main/java/org/apache/zookeeper/server/util/MessageTracker.java
+++ b/zookeeper-server/src/main/java/org/apache/zookeeper/server/util/MessageTracker.java
@@ -145,1 +145,1 @@
-            i = serverAddr.indexOf(':');
+            i = serverAddr.indexOf(':', i + 1);
\end{lstlisting}

\begin{lstlisting}[language=, caption={multi-agent-gpt5 patch}, label={lst:zk_gpt5_patch}]
diff --git a/zookeeper-server/src/main/java/org/apache/zookeeper/server/util/MessageTracker.java b/zookeeper-server/src/main/java/org/apache/zookeeper/server/util/MessageTracker.java
index 91952ca..094437b 100644
--- a/zookeeper-server/src/main/java/org/apache/zookeeper/server/util/MessageTracker.java
+++ b/zookeeper-server/src/main/java/org/apache/zookeeper/server/util/MessageTracker.java
@@ -140,9 +140,8 @@ public class MessageTracker {
             return 1;
         }
         int cnt = 1;
-        while (i > 0) {
+        while ((i = serverAddr.indexOf(':', i + 1)) != -1) {
             cnt++;
-            i = serverAddr.indexOf(':');
         }
         return cnt;
     }
\end{lstlisting}

\subsection{System Behavior Analysis}

In this section, we analyze agent interactions and tool usage as a proxy for reasoning and iteration behavior. 

\subsubsection{\bf Agent Interaction in Multi-Agent System}
Table \ref{tab:multiagent-call-counts} reveals a clear stratification in how the multi-agent architecture allocates its reasoning effort across tasks. For most tasks, the workflow collapses to a minimal single-pass pattern: one context setup, one root-cause analysis, one SWE patching step, and a single strategy/summary, resulting in a relatively low total of 5–8 calls. In contrast, more complex tasks such as \texttt{tk-delta-05}, \texttt{tk-delta-06}, and \texttt{zk-delta-01} trigger heavy iteration across all stages, with large increases in context retrieval, reflection, and SWE calls. Notably, these high-cost cases show a strong coupling between reflection and SWE invocation counts, indicating that failed or partial patches repeatedly route the system back into analysis–patch–reflect loops. Overall, the results suggest that while the multi-agent design is efficient on straightforward vulnerabilities, its cost scales sharply when tasks induce repeated reconsideration of root causes or patch strategies. The iteration depth, rather than the mere presence of multiple agents, highlights the dominant driver of overhead.

\begin{table}[h]
\centering
\begin{tabular}{l|cccccc|c}
\hline
Task & Ctx & Refl & RCA & SWE & Strat & Sum & Tot \\
\hline
cc-delta-01      & 1  & 0  & 1 & 1  & 1 & 1 & 5  \\
cc-delta-03      & 5  & 4  & 1 & 5  & 1 & 1 & 17 \\
cc-delta-04      & 1  & 0  & 1 & 1  & 1 & 1 & 5  \\
cc-delta-06      & 1  & 0  & 1 & 1  & 1 & 1 & 5  \\
cc-delta-07      & 2  & 1  & 1 & 2  & 1 & 1 & 8  \\
pdfbox-delta-01  & 1  & 0  & 1 & 1  & 1 & 1 & 5  \\
poi-delta-01     & 1  & 0  & 1 & 1  & 1 & 1 & 5  \\
poi-delta-02     & 1  & 0  & 1 & 1  & 1 & 1 & 5  \\
tk-delta-02      & 1  & 0  & 1 & 1  & 1 & 1 & 5  \\
tk-delta-03      & 1  & 0  & 1 & 1  & 1 & 1 & 5  \\
tk-delta-05      & 12 & 3  & 9 & 12 & 3 & 3 & 42 \\
tk-delta-06      & 12 & 9  & 3 & 12 & 3 & 3 & 42 \\
zk-delta-01      & 18 & 15 & 3 & 18 & 3 & 3 & 60 \\
\hline
\end{tabular}
\caption{Agent call counts per task (successfully patched) for the multi-agent architecture using GPT-5.}
\label{tab:multiagent-call-counts}
\end{table}

\begin{figure}[t]
  \centering
  \includegraphics[width=\columnwidth]{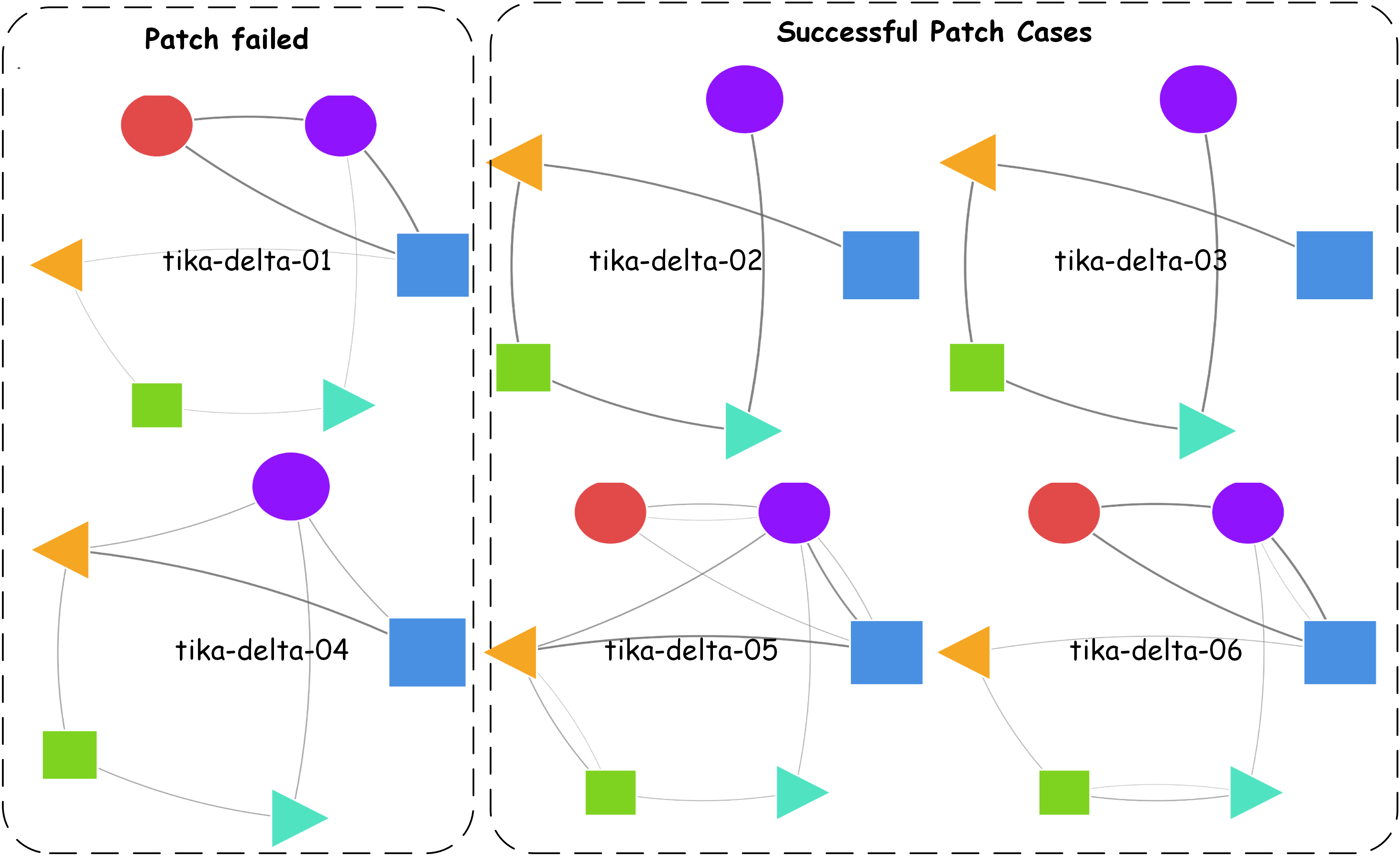}
  \caption{Agent-transition graph for Tika challenges.
  Nodes represent agents, and weighted edges represent the number of transitions between agents.}
  \label{fig:tika_agent_transition}
\end{figure}

Figure~\ref{fig:tika_agent_transition} shows the agent-transition graph of the multi-agent system for six Tika challenges. Nodes correspond to individual agents, and weighted edges indicate the frequency of transitions between agents. Among these tasks, \texttt{tk-delta-02}, \texttt{tk-delta-03}, \texttt{tk-delta-05}, and \texttt{tk-delta-06} were successfully patched, whereas \texttt{tk-delta-01} and \texttt{tk-delta-04} failed to past all tests, exhibiting distinct interaction patterns between successful and unsuccessful runs.

For simpler tasks such as \texttt{tk-delta-02} and \texttt{tk-delta-03}, each agent is typically invoked only once, and successful patches are produced without requiring reflection or iterative reasoning. The more challenging tasks \texttt{tk-delta-05} and \texttt{tk-delta-06} exhibit substantially richer interaction patterns, with frequent transitions among multiple agents. In these cases, many execution paths repeatedly route through the reflection agent, which then redirects control to other agents for subsequent reasoning and patch attempts. Across successful runs, the most frequently invoked agents are the context-retrieval agent, the SWE (strategy) agent, and the reflection agent, highlighting their central roles in patch refinement.

For failed tasks, the context-retrieval agent dominates the interaction graph. This suggests that the system repeatedly attempted to identify relevant code regions but was unable to converge on a satisfactory localization of the vulnerability, preventing downstream agents from effectively generating a correct patch.

\subsubsection{\bf Tool Usage in Single-Agent System}
Table \ref{tab:tool-call-counts} highlights clear differences in how the single-agent architecture interacts with code and validation tools across tasks. Overall, search and inspection operations dominate tool usage, particularly 'search\_read\_source' and 'search\_find\_references', indicating that most effort is spent understanding code structure. Edit actions ('editor\_apply\_change', 'editor\_undo\_last\_patch') are relatively infrequent, suggesting that the agent typically converges on a small number of concrete patch attempts once sufficient context is gathered. The number of 'test\_patch' calls remains low, implying that the single-agent workflow favors front-loaded analysis followed by limited validation iterations, rather than repeated patch–test cycles.

Failed tasks exhibit markedly different tool-usage patterns, characterized by excessive context retrieval and imbalanced editing behavior (see Appendix~\ref{app:tool-single-agent-failure}).

\begin{table*}[t]
\centering
\small
\setlength{\tabcolsep}{2.5pt}
\renewcommand{\arraystretch}{1.05}
\resizebox{\textwidth}{!}{%
\begin{tabular}{l|rrrrrrrrrrr|r}
\hline
Task &
\multicolumn{1}{c}{\makecell{editor\_\\apply\_\\change}} &
\multicolumn{1}{c}{\makecell{editor\_\\list\_\\edits}} &
\multicolumn{1}{c}{\makecell{editor\_\\undo\_\\last\_\\patch}} &
\multicolumn{1}{c}{\makecell{search\_\\find\_\\references}} &
\multicolumn{1}{c}{\makecell{search\_\\get\_\\callees}} &
\multicolumn{1}{c}{\makecell{search\_\\get\_\\callers}} &
\multicolumn{1}{c}{\makecell{search\_\\list\_\\functions}} &
\multicolumn{1}{c}{\makecell{search\_\\list\_\\types}} &
\multicolumn{1}{c}{\makecell{search\_\\read\_\\definition}} &
\multicolumn{1}{c}{\makecell{search\_\\read\_\\source}} &
\multicolumn{1}{c}{\makecell{test\_\\patch}} &
Total \\
\hline
cc-delta-01              &  2 & 0 & 0 &  6 & 0 & 0 & 2 & 4 & 0 &  22 &  2 &  38 \\
cc-delta-03              &  2 & 0 & 0 & 20 & 0 & 0 & 4 & 6 & 4 &  48 &  2 &  86 \\
cc-delta-04              &  1 & 0 & 0 &  2 & 0 & 0 & 0 & 4 & 0 &  16 &  2 &  25 \\
cc-delta-05              &  0 & 0 & 0 &  0 & 0 & 0 & 0 & 2 & 0 &   2 &  2 &   6 \\
cc-delta-06              &  2 & 0 & 0 &  0 & 0 & 2 & 2 & 2 & 0 &   4 &  2 &  14 \\
pdfbox-delta-01          &  0 & 0 & 0 &  0 & 0 & 0 & 0 & 2 & 0 &   0 &  2 &   4 \\
poi-delta-01             &  2 & 0 & 0 &  2 & 0 & 0 & 4 & 2 & 0 &  20 &  2 &  32 \\
tk-delta-02              & 18 & 2 & 0 &  2 & 0 & 0 & 0 & 4 & 4 &  58 &  2 &  90 \\
tk-delta-03              &  2 & 0 & 0 &  0 & 0 & 0 & 0 & 2 & 2 &   2 &  2 &  10 \\
tk-delta-04              &  6 & 0 & 0 &  0 & 0 & 0 & 4 & 6 & 2 &  44 &  2 &  64 \\
tk-delta-05              &  2 & 0 & 0 &  0 & 0 & 0 & 2 & 2 & 2 &   8 &  2 &  18 \\
zk-delta-01              &  2 & 0 & 0 &  2 & 0 & 0 & 4 & 2 & 2 &  50 &  2 &  64 \\
\hline
\end{tabular}%
}
\caption{Tool call counts per task (successfully patched) for single-agent architecture.}
\label{tab:tool-call-counts}
\end{table*}

\subsubsection{\bf Tool Usage in General Coding System}

The general coding agent exhibits a more exploratory and execution-driven tool usage pattern. As shown in Table~\ref{tab:tool-usage-per-task}, it relies heavily on \texttt{Bash} and \texttt{Read}.
Patch-specific systems similarly rely on context-retrieval tools for frequent code inspection. However, the general coding agent can autonomously write scripts to inspect and search code and execute them via \texttt{Bash}, providing substantially greater flexibility.

The use of \texttt{TodoWrite} further highlights the agent’s explicit planning behavior, enabling flexible reasoning over long patching trajectories. Such extensive exploration also incurs higher cost, revealing a clear trade-off between the adaptability of general coding agents and the efficiency of more structured patch-specific systems.

\subsection{Cost \& Latency}

Claude Code exhibits the highest token consumption, followed by the multi-agent architecture, with single-agent and fixed-workflow systems using comparable fewer tokens, as shown in Table \ref{tab:token-usage}. The high token usage of Claude Code can be attributed to its design as a general-purpose coding agent rather than a domain-specific patching system. Claude Code performs on-demand context retrieval, interactive reasoning, and tool-driven iteration across the entire project. Multi-agent systems often consume one to two orders of magnitude more tokens than single-agent setups. This behavior is expected: multi-agent designs repeatedly invoke root-cause analysis, planning, patch generation, and reflection cycles, each of which introduces additional LLM calls and accumulates long intermediate contexts. In contrast, the single-agent architecture exhibits a more direct loop with fewer reasoning steps in one iteration. The fixed workflow setup generally uses the fewest tokens, since its patch generation is tightly scoped by PoV traces and commit diffs.


Execution time does not correlate as strongly with token usage, as shown in Table \ref{tab:time-duration}. In several tasks, fixed workflow exhibits comparable or even longer runtimes than LLM-heavy approaches. This can be caused by environment setup or validation costs, such as building container images, running fuzzers, executing test harnesses. Architectures with extensive iteration incur build-time overhead that dominates LLM calls.
Claude Code does not hard-code exhaustive validation requirements. As a result, Claude Code tends to exhibit stable runtimes typically around 10 minutes per task. By comparison, other architectures often run until a predefined timeout for failed cases, and even successful runs may take 20–30 minutes due to repeated cycles. This highlights a fundamental trade-off between thorough validation and predictable execution time across different architectural designs.


\begin{table*}[t]
\centering
\small
\begin{tabular}{lcccccccc}
\hline
\textbf{Task} & \textbf{Bash} & \textbf{Read} & \textbf{Edit} & \textbf{TodoWrite} & \textbf{Write} & \textbf{Grep} & \textbf{Glob} & \textbf{Other} \\
\hline
cc-delta-01 & 17 & 7 & 2 & 7 & -- & -- & -- & -- \\
cc-delta-02 & 15 & 13 & 1 & 5 & -- & 7 & -- & -- \\
cc-delta-03 & 9 & 8 & 1 & 4 & -- & 1 & -- & -- \\
cc-delta-04 & 8 & 9 & 1 & -- & -- & -- & -- & -- \\
cc-delta-05 & 4 & 4 & 1 & -- & -- & -- & -- & -- \\
cc-delta-06 & 8 & 7 & 1 & -- & -- & -- & -- & -- \\
cc-delta-07 & 29 & 8 & 1 & -- & 2 & -- & -- & -- \\
logging-log4j2-delta-01 & 5 & 5 & 1 & 5 & -- & -- & 1 & -- \\
pdfbox-delta-01 & 7 & 7 & 2 & -- & -- & -- & -- & -- \\
poi-delta-01 & 4 & 6 & 1 & 4 & -- & 2 & -- & -- \\
poi-delta-02 & 11 & 5 & 1 & 7 & -- & -- & -- & -- \\
tk-delta-01 & 21 & 5 & 4 & 7 & -- & -- & -- & -- \\
tk-delta-02 & 6 & 6 & 2 & 4 & -- & -- & -- & -- \\
tk-delta-03 & 19 & 10 & 2 & -- & 3 & 1 & -- & -- \\
tk-delta-04 & 4 & 6 & 2 & 4 & 4 & -- & -- & -- \\
tk-delta-05 & 21 & 11 & 2 & 4 & -- & -- & 2 & -- \\
tk-delta-06 & 17 & 7 & 1 & 5 & 1 & -- & -- & 1 \\
zk-delta-01 & 8 & 6 & 1 & 8 & -- & -- & -- & -- \\
zk-delta-02 & 21 & 10 & 3 & 4 & 1 & 4 & -- & -- \\
\hline
\end{tabular}
\caption{Claude Code tool usage.}
\label{tab:tool-usage-per-task}
\end{table*}

\begin{table*}[t]
\centering
\scriptsize
\setlength{\tabcolsep}{3pt}
\resizebox{\textwidth}{!}{%
\begin{tabular}{l|ccccccccccccccccccc}
\hline
Mode 
& cc-01 & cc-02 & cc-03 & cc-04 & cc-05 & cc-06 & cc-07  
& pdfbox & poi-01 & poi-02 
& tk-01 & tk-02 & tk-03 & tk-04 & tk-05 & tk-06 
& zk-01  \\
\hline
SA-g 
& 9957 & x & 18218 & 6147 & 4979 & 3608 & x 
& 7646 & 71876 & x 
& x & 31247 & 10110 & 25168 & 6964 & x 
& 9317  \\
MA-g 
& 78138 & x & 254650 & 15254 & x & 34644 & 93256 
& 21855 & 91933 & 117294 
& x & 24723 & 24498 & x & 192696 & 476454 
& 585606  \\
FW-g  
& 20928 & 7690 & 3585 & x & x & x & 30033 
& 3349 & 16768 & x 
& 8935 & 5523 & 8218 & 9283 & 2905 & x 
& 2856  \\
SA-c 
& 57759 & x & x & 57702 & 33518 & 165894 & 2242242 
& 38917 & x & x & 945884 & 96004 & 2026827 & 587483 
& 88982 & 94554 & 33264 \\
MA-c 
& 111072 & x & x & 181032 & x & 138420 & x 
& 59574 & 1138527 & x & x & 74058 & x & x 
& x & x & 106251 \\
FW-c  
& x & x & 7982 & x & x & x & 19749 
& 39145 & 34019 & 79176 & 48037 & x & 23934 & 75323 
& 10458 & x & 114643 \\
Claude
& 216726 & 287732 & 176167 & 184960 & 179198 & 172931 & 319269 
& 133166 & 341188 & 366665 & 245091 & 102717 & 177962 & 133284 
& 157272 & x & x \\
\hline
\end{tabular}
}
\caption{Token usage across tasks. \texttt{x} denotes that the vulnerability was not patched.
Suffix \texttt{-c} denotes configurations using Claude Sonnet~4.5, \texttt{-g} denotes configurations
using GPT-5, and \texttt{Claude} corresponds to Claude Code.
Tasks for which fewer than two configurations successfully patched are excluded.}
\label{tab:token-usage}
\end{table*}

\begin{table*}[t]
\centering
\small
\resizebox{\textwidth}{!}{%
\begin{tabular}{l|ccccccccccccccccc}
\hline
Mode 
& cc-01 & cc-02 & cc-03 & cc-04 & cc-05 & cc-06 & cc-07 
& pdfbox-01 & poi-01 & poi-02 
& tk-01 & tk-02 & tk-03 & tk-04 & tk-05 & tk-06 
& zk-01 \\
\hline
SA-g 
& 343 & x & 429 & 301 & 306 & 300 & x 
& 301 & 554 & x 
& x & 1093 & 552 & 963 & 662 & x 
& 500  \\
MA-g
& 392 & x & 1767 & 457 & x & 631 & 1330 
& 426 & 695 & 740 
& x & 738 & 592 & x & 2313 & 2268 
& 3198 \\
FW-g 
& 680 & 1589 & 1595 & x & x & x & 4860 
& 278 & 687 & x 
& 1346 & 883 & 858 & 1700 & 881 & x 
& 728  \\
SA-c
& 317 & x & x & 287 & 286 & 296 & 2377 
& 268 & x & x & 669 & 516 & 780 & 1144 
& 569 & 820 & 328 \\
MA-c
& 362 & x & x & 464 & x & 468 & x 
& 334 & 1005.1 & x & x & 561 & x & x 
& x & x & 902 \\
FW-c
& x & x & 466 & x & x & x & 425 
& 816 & 492 & 576 & 1056 & x & 748 & 1376 
& 650 & x & 3172 \\
Claude
& 709 & 587 & 193 & 317 & 197 & 254 & 604 
& 357 & 326 & 416 & 622 & 170 & 682 & 282 
& 468 & x & x \\
\hline
\end{tabular}%
}
\caption{Execution time (seconds) across tasks. \texttt{x} denotes the vulnerability was not patched.}
\label{tab:time-duration}
\end{table*}




\input{related-work-qx}

\section{Conclusion}
We demonstrate that the effectiveness of LLM-based automated vulnerability repair depends critically on architectural design in addition to model capability. Our evaluation of four paradigms (namely fixed workflows, patch-specific agents, multi-agent systems, and general-purpose coding agents) on real-world AIxCC vulnerabilities reveals clear trade-offs in performance, robustness, and cost. No single architecture consistently dominates: general-purpose coding agents achieve broader coverage and robustness, while patch-specific agents offer greater efficiency and control. These findings establish architecture as a first-class consideration in the design of effective autonomous vulnerability repair systems.


\bibliographystyle{ACM-Reference-Format}
\bibliography{references}

\appendix
\section{Tool-Usage in Failed Single-Agent Patching}
\label{app:tool-single-agent-failure}
For patching tasks that were not successfully resolved by the single-agent system, as shown in Table~\ref{tab:failed-tool-usage}, failures are marked by heavy reliance on context-retrieval tools, particularly {\tt search\_read\_\\source} and {\tt search\_find\_references}. Editing actions ({\tt editor\_\\apply\_change}) are either sparse or unevenly distributed. Some failed tasks involve extensive exploration with minimal concrete edits (e.g., \texttt{tk-delta-01}), whereas others apply numerous edits without converging on a correct fix (e.g., \texttt{tk-delta-06}). Moreover, several tasks terminate with zero or few 'test\_patch' invocations, suggesting that the system never reached a point where it considered the generated patch sufficiently correct to call validation.

\begin{table}[t]
\centering
\small
\begin{tabular}{l|rrrrr}
\hline
Task 
& Edit 
& FindRef 
& ListFunc 
& ReadSrc 
& Test  \\
\hline
cc-delta-02 
& 52 & 8 & 4 & 114 & 6  \\
cc-delta-07 
& 36 & 4 & 2 & 150 & 0  \\
logging-log4j2-delta-01 
& 36 & 22 & 2 & 114 & 8  \\
poi-delta-02 
& 56 & 38 & 2 & 80 & 10  \\
tk-delta-01 
& 4 & 41 & 0 & 138 & 0  \\
tk-delta-06 
& 66 & 10 & 2 & 86 & 14  \\
\hline
\end{tabular}
\caption{Statistics of most frequently invoked tools for failed tasks in the single-agent system.}
\label{tab:failed-tool-usage}
\end{table}

\section{Ethical Considerations}

Our work focuses on analyzing and improving the architectural design of automated vulnerability patching systems, with the goal of strengthening software security and reducing the time between vulnerability discovery and remediation. The systems studied in this paper are evaluated exclusively in controlled environments using the AIxCC benchmark and sandboxed execution, and are not applied to production systems.

The AIxCC benchmark data are subject to access restrictions and are not publicly released by the challenge organizers. We respect these constraints and do not redistribute vulnerable programs, exploits, or Proofs of Vulnerability (PoVs). We do not release exploit generation components or any code that could be directly used to compromise real-world systems. All patching functionality is designed to support defensive purposes. The use of external LLM services follows the providers’ acceptable use policies, and no user data or personal information is collected or processed.

\section{Open Science Appendix}

This paper follows the ACM CCS Open Science policy and aims to support transparency, reproducibility, and long-term research impact. We provide an open-source artifact that implements a unified framework for comparing the architectures of automated patching systems based on Large Language Models (LLMs). The artifact includes three representative systems: \texttt{patch-agent-tools} (a tool-augmented single-agent system), \texttt{multi-agent} (a multi-agent system), and \texttt{patch-delta} (a fixed-workflow system). It further contains the unified runner script (\texttt{run\_patch\_system.py}), logging infrastructure, and documentation. Together, these artifacts enable reviewers to evaluate the architectural comparisons and analyses presented in the paper.

Our experiments are conducted on the AIxCC delta-scan Java vulnerability benchmark. However, the AIxCC benchmark data are not yet publicly open-sourced and are subject to access restrictions imposed by the challenge organizers. As a result, we cannot redistribute the original benchmark programs, vulnerabilities, or Proofs of Vulnerability (PoVs).

All artifacts are hosted in an anonymous repository and can be accessed by the program committee at:
\begin{quote}
\url{https://anonymous.4open.science/r/Patch-Architectures-07C6}
\end{quote}
The repository does not expose author identities, access logs, or tracking information. It includes detailed instructions for environment setup, dependency installation, system configuration, and reproduction of key experimental results. Our framework relies on external LLM services (e.g., OpenAI and Anthropic), which are accessed through standard APIs or via a LiteLLM proxy. 



\end{document}

%% file: arch.tex
\section{Four Architectural Paradigms}

In this section, we present four different architectures of LLM-based automated patching systems in detail, namely {\em Fixed-Workflow}, {\em Single-Agent}, {\em Multi-Agent} and {\em General-Code-Agent}.
To enable in-depth understanding and reproducibility, we provide a concrete and implementation-oriented characterization for each architecture.
\begin{figure*}[!t]
  \centering
  \includegraphics[width=0.85\textwidth]{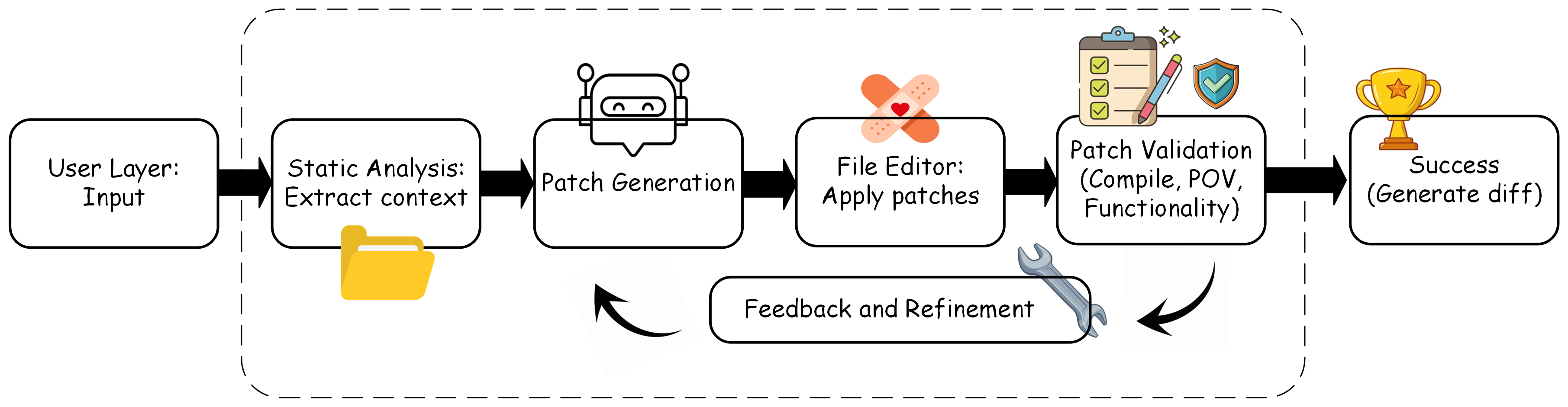}
  \caption{Fixed Workflow.}
  \label{fig:workflow}
\end{figure*}

\subsection{Fixed Workflow}

As illustrated in Figure~\ref{fig:workflow}, a fixed-workflow patching system follows a deterministic, rule-driven pipeline in which every patch attempt proceeds through a predefined sequence of steps. It does not have dynamic planning, but orchestrates a strictly ordered chain of components:
\begin{table}[h]
\centering
\small
\begin{tabular}{>{\centering\arraybackslash}p{2cm} p{5cm}}
\hline
\textbf{Stage} & \textbf{Description} \\
\hline
\textbf{User Layer} &
User provides source project, fuzzers, a commit diff, Proof-of-Vulnerability (PoV),  and crash information. \\

\textbf{Static Analysis} &
Analyze the code to locate suspicious functions with surrounding context. \\

\makecell{\textbf{Patch}\\\textbf{Generation}} &
Invoke the LLM with the constructed prompt to generate candidate patches. \\

\textbf{File Editor} &
Apply the generated patch to the codebase under a sandboxed execution environment. \\

\textbf{Patch Validation} &
Validate the patch through compilation, PoV tests, and functional checks. \\

\textbf{Feedback and Refinement} &
On failure, collect structured feedback and append it to the prompt for iterative refinement. \\
\hline
\end{tabular}
\caption{Components of the fixed patching workflow.}
\label{tab:patch-workflow}
\end{table}

Despite incorporating multi-turn dialogue and iterative correction, the overall fixed workflow architecture is static: the system does not reason about alternative strategies, reprioritize tasks, or reconfigure its workflow. Instead, patch generation is viewed as executing a fixed pipeline: identifying target functions to patch, generating candidate patches, validating them, and repeating until success or timeout. This design keeps the entire process transparent and easy to debug, and its structural simplicity makes adding or removing components straightforward.

Below, we describe each component of the patching workflow and its role in the overall patch synthesis process.



\subsubsection{\bf User Layer:}
The User Layer is the only interface through which users interact with the system. At this stage, users provide the source code of the target project together with auxiliary inputs required for automated patching. These inputs include a buggy commit that introduces the vulnerability, a Proof of Vulnerability (PoV) that specifies concrete inputs or execution conditions triggering the vulnerability, fuzzers that serve as the entry point for exercising and validating the PoV, and a set of functional tests that capture the expected program behavior.

\subsubsection{\bf Static Analysis:}
In fixed workflow systems, inputs flow sequentially from the user layer to a static analysis component that provides additional program context for later patch generation stage. The primary role of static analysis in an LLM-based patching system is to provide verifiable, structured evidence to the LLM, thereby reducing the search space during patch generation. Static analysis is crucial for identifying fault-relevant code regions, such as tracing how untrusted inputs propagate to dangerous sinks.

A common approach is to derive function names, or line numbers from crash reports or stack traces, and then extract the minimal relevant code snippets. Rather than emitting verbose logs, static analysis present its results to the LLM in effective formats, such as structured summaries including call paths, vulnerable function bodies, or recommended patch locations.

Typical tools used in this stage include CodeQL, Joern, SVF, and clang-analyzer. This stage does not require LLM involvement and is usually implemented as tools that can rapidly index and query the entire codebase. A key limitation, however, is that failures may be caused by global variables or shared state defined outside the functions in stack traces. In such cases, relying solely on call paths and crash stacks are insufficient, and broader program context should be considered.

\subsubsection{\bf LLM-Based Patch Generation:}
Patch generation is performed by invoking an LLM with a structured prompt integrating multiple sources of information. Prompt design is therefore a critical component of the system. In addition, explicit patching constraints are included to guide the model toward preserving existing functionality and limiting the scope of code changes. The goal of this stage is to organize all relevant information into a concise and task-focused representation for LLM to synthesize targeted patches. Listing~\ref{lst:patch_prompt} shows an example prompt template used for LLM-based patch generation.

Because patch generation often operates under strict context window constraints, effective context management is essential. In practice, context construction follows a prioritization strategy that allocates tokens preferentially to vulnerable code regions, static analysis results, and crash traces. When the assembled prompt exceeds the model’s context limit, truncation or summarization is applied using fixed heuristics that preserve semantic dependencies relevant to patch synthesis. Previously summarized context may also be cached and reused across iterations to reduce repeated token consumption.

To support fully automated application and validation, the LLM is required to emit patches in a predefined format, such as a unified diff or a replacement code snippet, without additional explanatory text. Outputs that do not conform to this format are discarded. Enforcing strict output constraints ensures that generated patches can be directly applied to the codebase without manual intervention.

\begin{lstlisting}[language=Python, caption={Example prompt template for LLM-based patch generation, where placeholders are populated with crash logs, static-analysis context and vulnerable functions.}, label={lst:patch_prompt}]
"""
# Vulnerability Patching Task

## Your Role
You are a world-leading security engineer tasked with fixing a vulnerability in code.
Your goal is to generate minimal, precise patches that address only the vulnerability without changing other functionality.
Do not apologize when you are wrong. Just keep optimizing the result directly and proceed with the process. Do not lie or guess when you are unsure about the answer.

## Input Information
### Vulnerability Stacktrace
{stacktrace}

### Context Information
The vulnerability is introduced by the following commit:
{commit_diff}

### Relevant Functions
{functions_metadata_str}

Please return the fixed functions to patch the vulnerability.

## Requirements
1. Fix ONLY the vulnerability--do not add features or refactor code
2. Preserve all existing functionality and logic
3. Make minimal changes (fewest lines of code possible)
4. Focus on security best practices

## Output Format
Return ONLY a JSON dictionary where keys are function names and values are code blocks:
{
  "function_name1": "function_content_with_fix",
  "function_name2": "function_content_with_fix"
}

IMPORTANT:
- Return the fixed content for each changed function
- Do NOT return diffs or partial code snippets
- Do NOT include explanations or comments outside the JSON
- Include ALL lines of the original function in your response, with your fixes applied

Return ONLY the JSON dictionary described above.
"""
\end{lstlisting}

Model invocation is typically implemented through configurable LLM abstraction or routing layers that decouple patching logic from specific model providers. Such layers are implemented using tools such as LiteLLM or custom lightweight clients, providing a unified interface over heterogeneous backends (e.g., OpenAI, Anthropic, Google, or local models). To improve robustness and throughput, routing strategies such as round-robin or priority-based selection are employed to distribute requests across multiple models, mitigating rate limits and transient failures. 
Additional practical engineering considerations include reducing latency and cost for high-frequency API calls through batching, response caching, and rate-limiting mechanisms.

\subsubsection{\bf File Editor:}
Different patching systems may adopt different file editing strategies. For example, an LLM may be instructed to rewrite one or more entire functions, replace specific lines (which requires high line-number accuracy and often multiple fuzzy-matching attempts), substitute old code snippets with new ones, or directly generate a unified diff file for a target source file. Prior work~\cite{afc-crs-all-you-need-is-a-fuzzing-brain} suggests that having LLMs generate diffs directly can lead to lower patch quality compared with generating new code and computing diffs afterward, potentially due to limitations in LLM reasoning capabilities.

Regardless of the editing strategy, it is essential to use a sandboxed environment to test whether patches can be applied cleanly, whether the modified code compiles, and whether it preserves intended functionality. Directly editing the source code risks permanently losing critical functionality. To address this, RoboDuck~\cite{aixcc-afc-archive} introduce a virtual file system (VFS) layer that records each edit. This design allows patches to be applied incrementally, reverted if necessary, and combined flexibly, greatly increasing robustness during iterative repair.

\subsubsection{\bf Patch Validation:}
A patch is considered valid only if it satisfies all correctness and security requirements. Specifically, the patched code must compile successfully and preserve the intended functionality of the program. In addition, the Proof of Vulnerability (PoV) is re-executed to ensure that the crash or exploit can no longer be triggered. All validation is conducted in a sandboxed environment, and any patch that fails compilation, functional tests, or PoV checks is discarded. If validation passes, a diff is generated as the final patch to be submitted.

\subsubsection{\bf Feedback and Refinement:}
In a fixed-workflow patching system, a failed patch typically triggers a subsequent iteration of the same workflow. A common strategy is to feed the failed patch and its associated execution logs back into the prompt for the next LLM invocation. This feedback enables the LLM to reason about the causes of failure, such as whether it expanded the attack surface by introducing new sinks or parsing paths, or whether it broke critical semantics by overly rejecting valid inputs.

By explicitly surfacing such failure signals, the feedback mechanism allows the patching system to revise vulnerable regions that may introduce new errors and progressively converge toward a correct and secure patch.

\begin{figure*}[!t]
  \centering
  \includegraphics[width=0.85\textwidth]{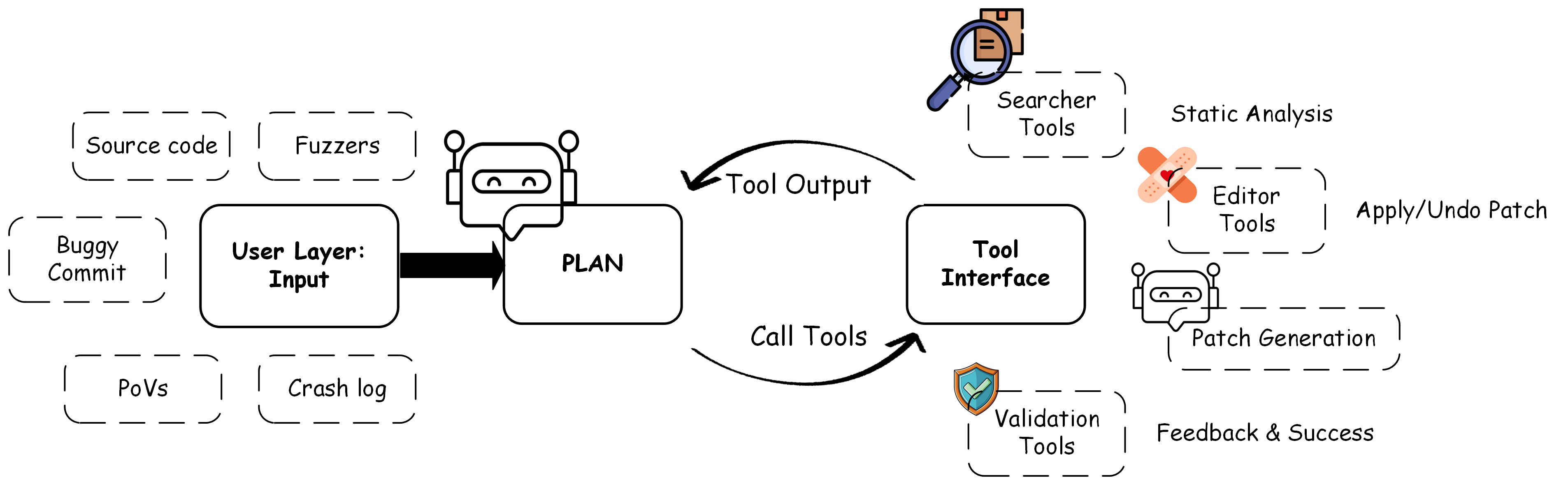}
  \caption{Single-Agent Patching System.}
  \label{fig:single-agent}
\end{figure*}

\subsection{Single-Agent System}
As illustrated in Figure~\ref{fig:single-agent}, a single-agent architecture equips an LLM with autonomy for iterative planning and structured tool invocation. Under this design, the agent conducts high-level reasoning to selects appropriate actions, such as searching, editing, generating, or validating code. When a generated fix fails to remediate the vulnerability, the agent incorporates feedback from failed tests into its subsequent planning step and uses it to guide the next action selection. This architecture highlights the core strength of single-agent tool use: the model independently reasons about the repair task, invokes tools to inspect and modify code, and iteratively validates its outputs with a structured validation tool.

\begin{table}[h]
\centering
\small
\begin{tabular}{>{\centering\arraybackslash}p{2cm} p{5cm}}
\hline
\textbf{Stage} & \textbf{Description} \\
\hline
\textbf{User Layer} &
User provide source project, fuzzers, a commit diff, Proof-of-Vulnerability (PoV),  and crash information. \\

\textbf{Planner} &
Model autonomously reasons about the patching task, invokes tools to inspect, modify and validate the code. \\

\textbf{Tool Interface} &
Operate over a suite of fine-grained code-interaction tools. \\
\hline

\end{tabular}
\caption{Components of the single-agent system.}
\label{tab:single-agent}
\end{table}


Several components in fixed-workflow systems also appear in single-agent architectures, but they are realized implicitly through tool use rather than explicit orchestration. Examples include patch application and validation. Consequently, we focus on the structural characteristics that distinguish single-agent patching systems from alternative architectures, rather than providing an exhaustive inventory of individual tools.

\subsubsection{\bf Planning and Reasoning:}
Planning and reasoning form the core of the single-agent architecture. A single LLM agent is responsible for interpreting proof-of-vulnerability (POV) descriptions, crash reports, and fuzzer code, and for deriving an overall patching strategy. Based on this reasoning, the agent dynamically selects and invokes the appropriate tool at each stage of the patching process.

Planning in this setting is implicit: there is no explicit phase separation, such as a predefined sequence of static analysis followed by patch generation. In single-agent architectures, planning is internalized within the model's chain-of-thought and expressed only through its tool-selection behavior. Code search, modification, and verification are interleaved within a continuous reasoning context. Although subagents may be introduced for tasks such as context retrieval and management, the core reasoning process remains unified within a single decision-making loop. 


\subsubsection{\bf Tool Interface:}

Each tool is explicitly registered with the agent, so that the agent is aware of the full set of actions it can perform. This tool interface is essential for enabling a single-agent system to operate analogously to a human developer interacting with a codebase. Listing~\ref{lst:tool_only_single_agent_prompt} illustrates how single-agent patching systems constrain the LLM to operate exclusively through structured tool invocation.

A single-agent patching system exposes a collection of fine-grained tools to the LLM, including:
\begin{itemize}
    \item \textbf{Searcher Tools:} SearchReadSource, SearchReadDefinition, and SearchFindReferences;
    \item \textbf{Editor Tools:} EditorApplyPatch, EditorUndoLastPatch, and EditorListEdits;
    \item \textbf{Patch Generation:} GeneratePatch;
    \item \textbf{Validation Tools:} TestPatch;
\end{itemize}

Editor tools support applying and reverting code changes (\texttt{Editor\\ApplyPatch}, \texttt{EditorUndoLastPatch}) as well as inspecting the current edit state (\texttt{EditorListEdits}). Searcher tools enable navigation and inspection of the codebase, including reference lookup and source or definition retrieval (\texttt{SearchFindReferences}, \texttt{SearchRead\\Definition}, \texttt{SearchReadSource}). Patch generation is performed by invoking the LLM through the \texttt{GeneratePatch} tool, while validation is carried out by executing the modified program via \texttt{TestPatch}.


\begin{lstlisting}[language=, caption={Tool-augmented prompt for a single-agent patching system}, label={lst:tool_only_single_agent_prompt}]
# SYSTEM (Tool-Augmented Agent)
You are responsible for fixing vulnerability by editing source code.
The fix must preserve intended behavior and be validated with PoVs.

Rules:
- Only modify <LANG> source files under <SOURCE_ROOT>
- Do NOT modify the fuzzing harness
- Fix the root cause of the vulnerability

# USER (Task Context)
<pov>
  <harness><FUZZER_NAME></harness>
  <fuzzer_data>... fuzzing / PoV inputs ...</fuzzer_data>
</pov>

# TOOL-DRIVEN INTERACTION
# The agent must interact with the codebase exclusively via tools.

Use SearchReadSource, SearchReadDefinition, and SearchFindReferences to inspect the code.
Use GeneratePatch to generate patches.
Use EditorApplyPatch to modify files (absolute path required) and EditorUndoLastPatch to revert edits.
Use EditorListEdits to review accumulated changes.
Use TestPatch to validate fixes; if validation fails, revise and retry.
When the patch is correct, call terminate.
\end{lstlisting}


All tool interactions are mediated through a standardized interface: each tool's JSON schema is embedded in the model prompt, the agent emits structured tool calls with explicit arguments, and the runtime executes the corresponding functions and returns results as tool messages. This interaction is commonly realized using mechanisms called the Model Context Protocol (MCP)~\cite{anthropic2024mcp}, which provides a standardized interface for integrating external tools and execution environments with LLMs. This design enables the agent to continuously observe changes to the code state, update its internal reasoning, and select subsequent actions based on tool feedback.

Listing~\ref{lst:editor_apply_change_schema} illustrates how an editing tool is exposed to a single-agent system through a structured JSON schema. This interface functions as a formal contract between the LLM and the execution environment, translating the model's natural-language reasoning into concrete, executable actions. 
By enforcing a structured, machine-readable contract, MCP decouples high-level reasoning from execution, enabling precise tool invocation without embedding execution logic in the prompt.

\begin{lstlisting}[language=, caption={Example JSON schema for an editing tool}, label={lst:editor_apply_change_schema}]
{
  "type": "function",
  "function": {
    "name": "EditorApplyPatch",
    "description": "Apply a change to 'path' using ONE of two modes:
    - Unified diff mode: provide 'patch' as a unified diff (with @@ hunks).
    - Snippet mode: provide 'old_code' and 'new_code'; the tool computes a diff and applies it.
    The applied change is tracked and can be undone via editor_undo_last_patch.",
    "parameters": {
      "type": "object",
      "properties": {
        "path": {
          "type": "string",
          "description": "Absolute path of the file to modify"
        },
        "patch": {
          "type": "string",
          "description": "Unified diff to apply (optional)"
        },
        "old_code": {
          "type": "string",
          "description": "Original code snippet to be replaced (optional)"
        },
        "new_code": {
          "type": "string",
          "description": "Replacement code snippet (optional)"
        },
        "replace_all": {
          "type": "boolean",
          "description": "Whether to replace all occurrences of old_code"
        }
      },
      "required": ["path"]
    }
  }
}
\end{lstlisting}


\subsubsection{\bf Feedback Loop:}
The feedback mechanism in a single-agent system differs fundamentally from that of workflow-based systems. In workflow-based designs, a failed patch typically triggers a full restart of the predefined workflow. In contrast, for a single-agent system, a failure is treated as an intermediate observation within the agent's ongoing reasoning trajectory rather than a terminal outcome.

After each patch attempt, the agent decides whether to invoke validation tools to run tests and collect failure signals such as crash logs. When a patch fails, the agent incorporates this feedback directly into its context and continues reasoning, refining the patch in subsequent steps. This forms an iterative feedback loop in which reasoning, code modification, and validation are tightly interleaved.

\subsection{Multi-Agent System}

\begin{figure}[h]
  \centering
  \includegraphics[width=\columnwidth]{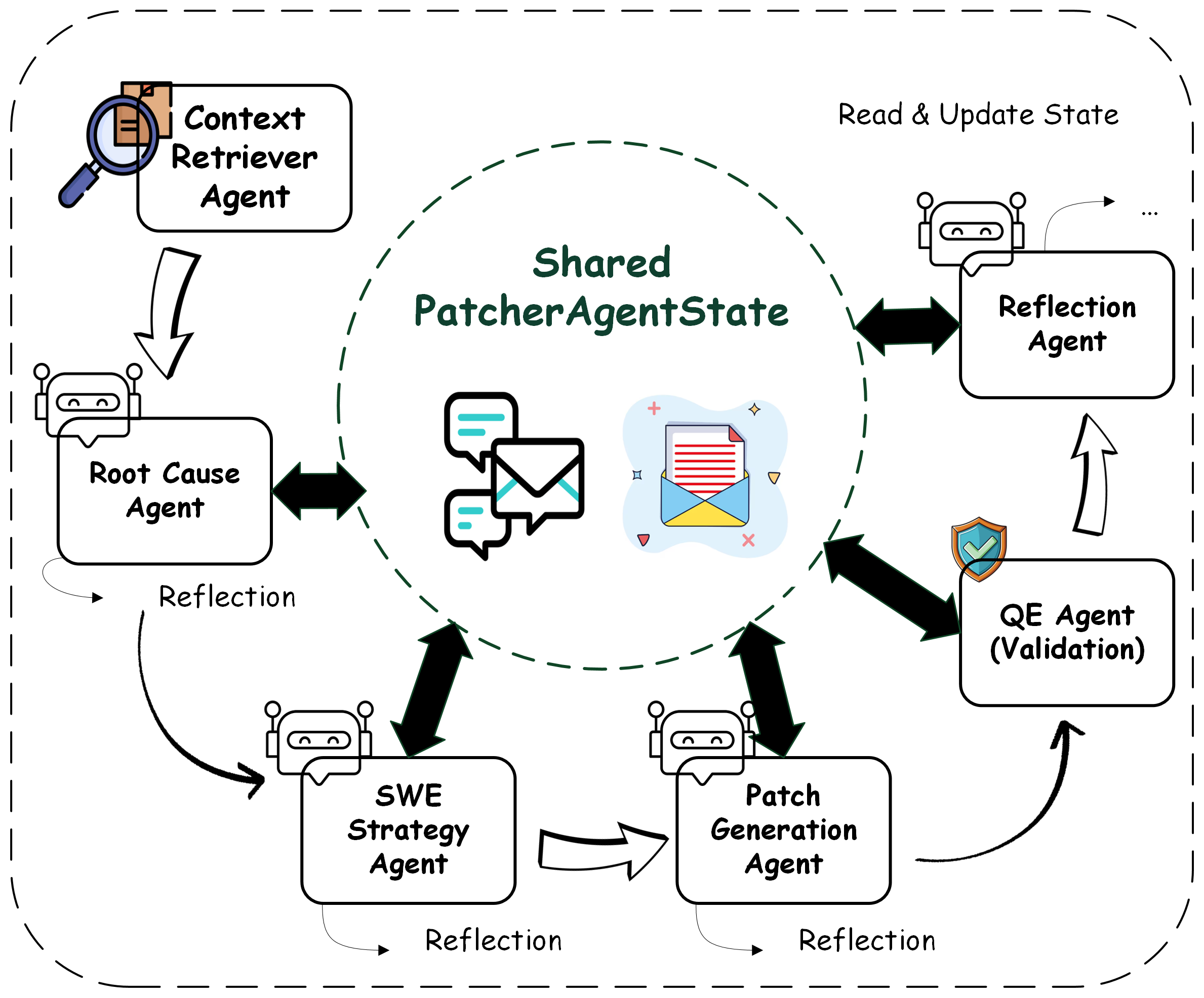}
  \caption{Multi-Agent Patching System.}
  \label{fig:multi-agent}
\end{figure}

As illustrated in Figure~\ref{fig:multi-agent}, multi-agent patching architectures decompose the repair workflow into specialized agents, each responsible for a distinct reasoning or execution function. This design addresses two key challenges in automated patching: the need for distinct reasoning modes (e.g., localization, synthesis, and validation) and the risk of context-window overload caused by long execution traces and large codebases. Assigning clear responsibilities to specialized agents reduces per-agent cognitive load and enables modular decision-making.


\begin{table}[h]
\centering
\small
\begin{tabular}{>{\centering\arraybackslash}p{2cm} p{5cm}}
\hline
\textbf{Stage} & \textbf{Description} \\
\hline
\textbf{User Layer} &
User provide source project, fuzzers, a commit diff, Proof-of-Vulnerability (PoV),  and crash information. \\

\textbf{Shared State} &
A shared state maintains all task-level information, including user input, relevant code snippets, root cause analysis, patch strategies, patch attempts, and message history. It also stores execution paths and \texttt{next\_agent} field to supports dynamic routing. \\

\textbf{Agents} &
Specialized agents perform distinct tasks such as context retrieval, patch generation, validation, and reflection.  \\
\hline
\end{tabular}
\caption{Components of the multi-agent system.}
\label{tab:multi-agent}
\end{table}


Rather than relying on a single agent to perform root cause analysis, context retrieval, patch synthesis, and validation, the system coordinates these steps through a shared state. Most agents operate directly on the shared message context, extending a unified conversation history, while some maintain local state during execution and synchronize their outputs with the global state upon completion. For example, the context-retriever agent maintains a private {\tt CodeSnippetManagerState} and populates relevant code snippets into the shared state after querying extraction tools for functions, types, and call relationships. This isolation prevents the global state from accumulating intermediate reasoning and low-utility artifacts.

Next, we describe the key structural components of multi-agent patching architectures in detail.

\subsubsection{\bf Shared State:}
A defining feature of the architecture is a shared state object that serves as the coordination substrate across agents. The shared state records the evolving patching session, typically including the conversation history, extracted code snippets, candidate patches, build/test outcomes, and reflection. Agents do not maintain long-lived private contexts; instead, each agent reads from and appends to a shared message buffer, enabling consistent memory across the pipeline.

Concretely, all agents operate over {\tt PatcherAgentState.messages}: when an agent is invoked, it resumes from the same shared message history, produces its output, and extends the shared state. 

\subsubsection{\bf Specialized Agents:}

This section categorizes agent roles by responsibility. The number of agents and their task partitioning significantly affect patching performance. Prior work shows that agents for property analysis, root-cause reasoning, strategy generation, and reflection improve patch correctness and quality~\cite{kim2025atlantis,afc-buttercup}.

\begin{itemize}
    \item \textbf{Context Retriever Agent:} The Context Retriever Agent is responsible for retrieving precise program context with minimal prompt overhead. It transforms crash stack traces into structured code queries, such as function/type requests and call-graph navigation requests. It then invokes extraction tools (e.g., get\_function, get\_type, get\_callers, get\_callees) to retrieve the most relevant code snippets and assemble a compact context package. By centralizing context acquisition, the system avoids repeatedly prompting patching agents with large, redundant code regions and reduces the risk of missing cross-cutting definitions (e.g., helper functions, type aliases, or indirect callers).

    \item \textbf{Root-Cause Agent:} Given the stack traces, code diffs (i.e., committed changes), and collected snippets, the Root-Cause agent hypothesizes the fault mechanism and localizes the likely vulnerable location, as illustrated in Listing~\ref{lst:prompt_templates}(1). Rather than immediately producing code edits, it outputs actionable guidance—which invariants appear violated, which input fields are untrusted, and where validation or sanitization should occur. This separation is critical: it prevents premature patch synthesis based on incomplete evidence and stabilizes downstream patch generation by grounding it in an explicit diagnosis.

    \item \textbf{SWE (Patch Strategy) Agent:}
    The SWE agent drafts a high-level patch strategy in natural language, typically describing the intended changes. It then summarizes the strategy into an implementation-ready plan that can be consumed by the patching stage. Structurally, this agent functions as a bridge between diagnosis and code generation: it converts root-cause hypotheses into concrete editing objectives and ensures that patch generation is guided by explicit design intent.

    \item \textbf{Patch Agent:}
    The Patch agent is responsible for producing a patch that can be applied reliably to the codebase, as illustrated in Listing~\ref{lst:prompt_templates}(2). To mitigate brittle edits, it requests old/new code with surrounding lines to support safe replacement and then emits a unified diff. This role separation is important in practice: forcing the same agent to both reason about the patch and produce application-stable diffs often increases formatting and alignment errors~\cite{kim2025atlantis}. 

    \item \textbf{QE (Validation) Agent:}
    The QE agent validates candidate patches by rebuilding the project, running PoVs to confirm remediation of the crash behavior, and executing regression tests. It produces structured outcomes (build success, crash status, failing tests) that are fed back into the shared state. 

    \item \textbf{Reflection Agent:}
    When validation fails, the Reflection agent analyzes the failure signals and provides targeted feedback to earlier stages. This includes diagnosing whether the patch missed the root cause, introduced a new sink/path, or broke key semantics. Instead of naively repeating the entire pipeline, reflection produces focused guidance that improves the next iteration’s efficiency and reduces oscillation across patch attempts.

\end{itemize}

\begin{lstlisting}[language=Python, caption={Prompt templates for RCA and patch agents}, label={lst:prompt_templates}]
# (1) Root-Cause Analysis (RCA) prompt
SYSTEM: You are a root-cause analysis agent. Perform a precise root-cause analysis for the vulnerability.
USER:
PROJECT: {PROJECT_NAME}
DIFF CONTEXT (truncated): {DIFF_CONTEXT}
RELEVANT CODE SNIPPETS: {CODE_SNIPPETS}
OUTPUT ONLY:
<root_cause> ...technical explanation tied to exact code logic/locations... </root_cause>

# (2) Patch-generation prompt 
SYSTEM:
You are a skilled software engineer generating minimal patch changes.
Return ONLY <patch> blocks in the required format; no prose.
CONSTRAINTS:
- Only modify files under the project source directory (SRC).
- Do NOT modify oss-fuzz / infra / pov / harness directories.
- <file_path> must be relative to the source root.

USER:
PROJECT: {PROJECT_NAME}
ROOT CAUSE (if available): {ROOT_CAUSE}
REFLECTION (if available): {REFLECTION_GUIDANCE}
PATCH STRATEGY (optional): {PATCH_STRATEGY}
EDITABLE SNIPPETS (edit ONLY these regions): {CODE_SNIPPETS}

OUTPUT FORMAT (one or more blocks):
<patch>
  <file_path>relative/path/from/source/root</file_path>
  <identifier>function_or_method_name</identifier>
  <old_code> ...exact region to replace with context... </old_code>
  <new_code> ...same region with fix applied... </new_code>
</patch>

\end{lstlisting}

\subsubsection{\bf Control Flow:}

Agent transitions in a multi-agent patching system are partially structured but not fully hard-coded. Rather than enforcing a single linear execution order, the system defines a set of permissible transitions between agent roles, allowing control flow to adapt dynamically based on intermediate outcomes and shared state. This design prevents uncontrolled agent invocation while still enabling iterative reasoning and recovery from failures.

In practice, the architecture specifies a core progress path: context retrieval → root-cause analysis → patch strategy → patch generation → validation. However, transitions are not strictly one-way. agents may redirect execution to a Reflection agent upon detecting inconsistencies or failures. Reflection serves as a convergence point reachable from multiple stages, rather than a fixed pipeline step. Instead of restarting the workflow, it analyzes signals in the shared state and produces targeted guidance that shapes subsequent transitions, enabling repair trajectories that are difficult to express in strictly sequential workflows.

\subsection{General Code Agent}

\begin{figure}[t]
  \centering
  \includegraphics[width=\columnwidth]{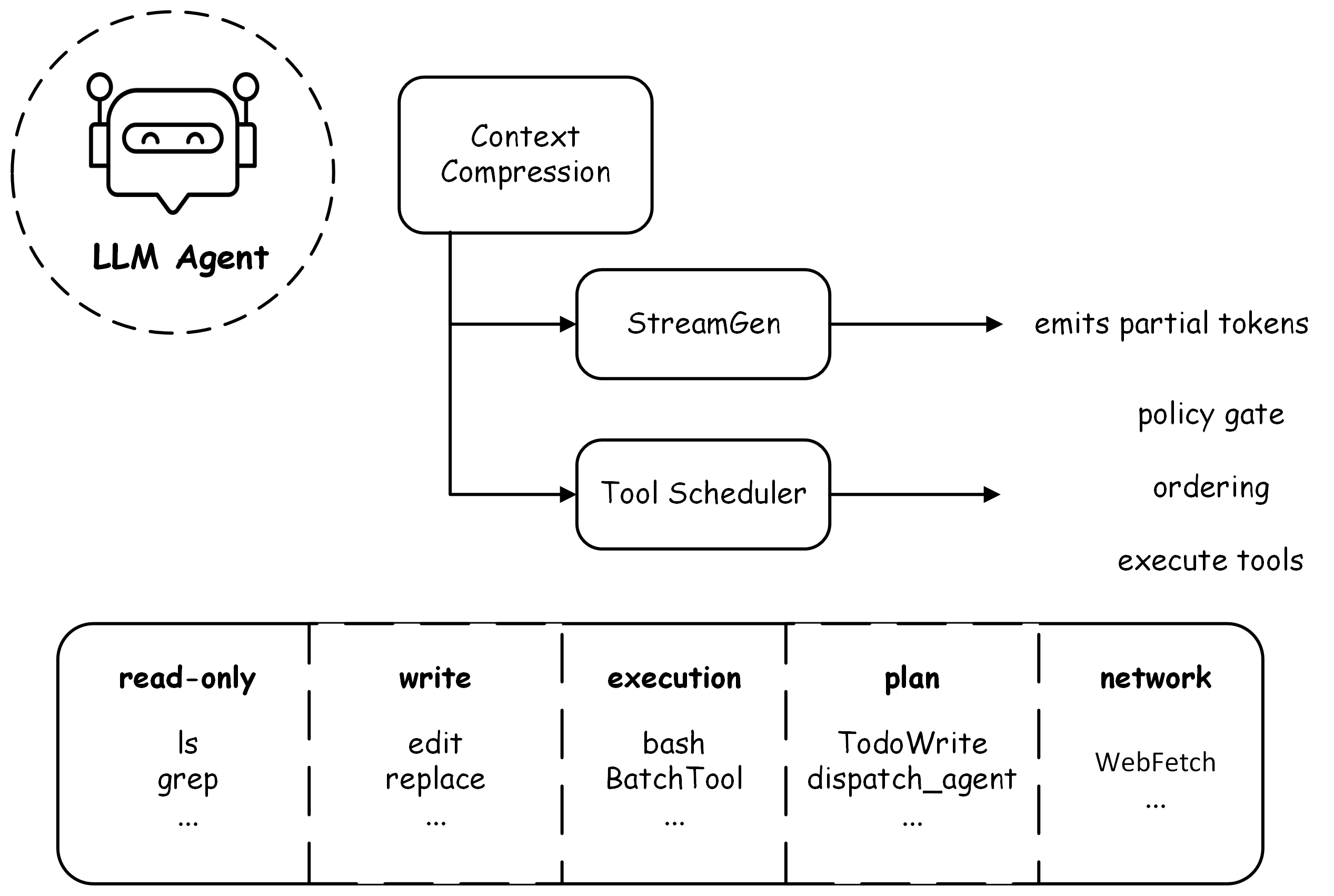}
  \caption{Architecture of a general coding agent.}
  \label{fig:general_agent_arch}
\end{figure}

General code agents, exemplified by Claude Code~\cite{anthropic_claudecode_2025}, represent a distinct class of AI systems that function as end-to-end coding assistants with internal reasoning, dynamic planning, and access to rich environment-manipulation tools. As illustrated in Figure~\ref{fig:general_agent_arch}, unlike domain-specific patching agents, these systems are not specialized for a single software task but instead act as developer copilots capable of reasoning over arbitrary projects and objectives. This generality enables capabilities beyond patching, including autonomous codebase exploration, context retrieval, code modification, and test execution.

\begin{table}[h]
\centering
\small
\begin{tabular}{>{\centering\arraybackslash}p{2cm} p{5cm}}
\hline
\textbf{Stage} & \textbf{Description} \\
\hline
\textbf{User Layer} &
User provide source project, fuzzers, a commit diff, Proof-of-Vulnerability (PoV), crash information and a \texttt{CLAUDE.md} file. \\

\textbf{Context Compression} &
Constructs a compact task context by selecting and summarizing relevant code regions and prior logs. \\

\textbf{StreamGen} &
Produces intermediate reasoning on the agent’s internal decision process, while remaining decoupled from tool invocation. \\

\textbf{Tool Scheduler} &
Orchestrates general-purpose tool calls to inspect the codebase, apply changes, and execute validation commands. \\

\hline
\end{tabular}
\caption{Components of the general coding system.}
\label{tab:general-agent}
\end{table}

\subsubsection{\bf Context Compression:}
The workflow begins when the LLM agent receives a task along with the current project context. Before each reasoning step, a context compression stage selectively summarizes prior interactions, recent tool outputs, and project-level instructions to fit within the model’s context window. This step ensures that long-running sessions remain tractable while preserving task-relevant state.

\subsubsection{\bf StreamGen:}
The compressed context is then fed into two parallel components. StreamGen is responsible for producing partial, user-facing outputs, incremental tokens rendered in the terminal, as the model generates them. Importantly, StreamGen is decoupled from the agent’s internal decision-making: it exposes intermediate text without committing to final actions, enabling responsive interaction without blocking on tool execution.

\subsubsection{\bf Tool Scheduler:}
In parallel, the agent’s structured outputs are routed to a tool scheduler, which interprets model-emitted tool calls and governs their execution. The scheduler enforces a policy gate that validates each requested action against safety and permission constraints, e.g., restricting filesystem writes, or controlling network access. Once approved, the scheduler imposes a well-defined ordering over tool invocations, ensuring that dependent actions (such as editing code before running tests) are executed sequentially.

Architecturally, most intelligence and decision-making reside within the model, which orchestrates tool use and interprets execution feedback, while the client-side runtime remains intentionally minimal and limited to tool execution. This separation avoids heavy scaffolding, supports rapid iteration, and enables rich IDE and subagent integrations. As a result, general code agents provide a flexible and extensible alternative to domain-specific patching workflows.

%% file: related-work-qx.tex
\section{Related Work}\label{sec:relatedwork}

\subsection{Automated Program Repair}
Early APR research explored rule-based, template-based, constraint-based, and search-based techniques, which rely on program analysis and predefined repair patterns to synthesize candidate patches~\cite{huang2024evolving}. 

The majority of these APR tools are test-driven. GenProg~\cite{6035728} applies genetic programming, and SemFix~\cite{6606623} leverages symbolic execution. Angelix~\cite{10.1145/2884781.2884807} scales semantic repair through a set of searches over angelic values. Prophet~\cite{10.1145/2837614.2837617} trains probabilistic models to prioritize candidate patches. ACS~\cite{xiong2017precise} improves repair precision by leveraging dependency analysis, API documentation, and mined predicates, while SimFix~\cite{jiang2018shaping} searches for correct patches by reusing existing code and historical fixes. Although effective for certain classes of bugs, these approaches are often limited by restricted patch spaces and brittle heuristics.

More recent learning-based methods address these limitations by directly predicting patches, often framing repair as a translation or generation task~\cite{zhang2023survey, xia2023automated}. These approaches demonstrate clear scaling effects, with larger models achieving better performance and patch diversity, but still rely on carefully designed pipelines for localization, context selection, and validation.

\subsection{LLM-Based Program Repair}

Early LLM-based repair work focused on prompt engineering and structured context construction, including chain-of-thought–guided prompting~\cite{ahmed2023better} and semantics-aware pipelines~\cite{nong2025appatch}. AutoCodeRover~\cite{zhang2024autocoderover} follows a largely fixed workflow in which fault localization and retrieval are performed upfront, and the LLM is invoked only for patch generation within a constrained context. These methods improve patch quality but typically embed LLMs within fixed workflows, limiting their ability to adaptively reason the repair process.

Several recent systems move beyond static pipelines. OpenHands~\cite{wang2024openhands} and SWE-agent~\cite{yang2024swe} provide a generalist agent framework that supports autonomous planning, tool usage, and multi-agent coordination, enabling more flexible repair strategies. However, these systems differ substantially in how control flow, reasoning, and tool interaction are structured, making their architectural trade-offs difficult to assess.

\subsection{Agent-Based and Tool-Augmented Architectures}

Recent APR research increasingly adopts agent-based designs, where LLMs are equipped with explicit reasoning, planning, and tool-use capabilities. Tool-augmented LLMs extend models beyond text generation by enabling interaction with external tools such as compilers, analyzers, retrieval systems and test frameworks. Foundational work such as ToolLLM~\cite{qin2023toolllm}, Toolformer~\cite{schick2023toolformer}, and Gorilla~\cite{patil2024gorilla} demonstrates how LLMs can learn to invoke tools reliably, and these ideas have directly influenced APR systems. For example, Agentless~\cite{xia2025demystifying} leverages execution, testing, and validation tools within a lightweight agentic loop, achieving strong performance on SWE-bench.

Multi-agent architectures further decompose the repair process by assigning specialized roles to different agents, such as localization, patch generation, and verification. Systems such as FixAgent (UniDebugger)~\cite{lee2024unified} and VarPatch~\cite{zhang2024poster} show that agent collaboration and conversational prompting can improve repair coverage and expressiveness. At the same time, recent analyses highlight new risks: red-teaming studies such as SWExploit show that multi-agent systems can generate patches that appear correct but remain vulnerable, raising concerns about robustness and safety~\cite{chen2025red}.

\subsection{General-Purpose Coding Agents}

In parallel, general-purpose coding agents such as Claude Code, GitHub Copilot, and Codex have demonstrated strong performance on real-world programming tasks, such as vulnerability remediation~\cite{anthropic_claudecode_2025, github_copilot_2025, chen2021codex}. These agents typically operate in an iterative loop, alternating between reasoning over code, invoking tools, and applying edits. Although not designed specifically for APR benchmarks or security competitions, they represent a distinct architectural paradigm that emphasizes broad autonomy and adaptability over task-specific structure.

%% file: references.bib
@misc{log4shell,
  title        = {CVE-2021-44228: Apache Log4j2 Remote Code Execution Vulnerability (Log4Shell)},
  howpublished = {\url{https://nvd.nist.gov/vuln/detail/CVE-2021-44228}},
  year         = {2021},
  note         = {Accessed: 2026-01},
  organization = {National Vulnerability Database},
  doi          = {10.1000/nvd-cve-2021-44228}
}

@misc{mongobleed,
  title        = {CVE-2025-14847: MongoDB Unauthenticated Heap Memory Disclosure Vulnerability (MongoBleed)},
  howpublished = {\url{https://nvd.nist.gov/vuln/detail/CVE-2025-14847}},
  year         = {2025},
  note         = {Accessed: 2026-01},
  organization = {National Vulnerability Database},
  doi          = {10.1000/nvd-cve-2025-14847}
}

@inproceedings{ahmed2023better,
  title={Better patching using llm prompting, via self-consistency},
  author={Ahmed, Toufique and Devanbu, Premkumar},
  booktitle={2023 38th IEEE/ACM International Conference on Automated Software Engineering (ASE)},
  pages={1742--1746},
  year={2023},
  organization={IEEE}
}

@inproceedings{nong2025appatch,
  title={$\{$APPATCH$\}$: Automated adaptive prompting large language models for $\{$real-world$\}$ software vulnerability patching},
  author={Nong, Yu and Yang, Haoran and Cheng, Long and Hu, Hongxin and Cai, Haipeng},
  booktitle={34th USENIX Security Symposium (USENIX Security 25)},
  pages={4481--4500},
  year={2025}
}

@article{wang2024openhands,
  title={Openhands: An open platform for ai software developers as generalist agents},
  author={Wang, Xingyao and Li, Boxuan and Song, Yufan and Xu, Frank F and Tang, Xiangru and Zhuge, Mingchen and Pan, Jiayi and Song, Yueqi and Li, Bowen and Singh, Jaskirat and others},
  journal={arXiv preprint arXiv:2407.16741},
  year={2024}
}

@inproceedings{zhang2024autocoderover,
  title={Autocoderover: Autonomous program improvement},
  author={Zhang, Yuntong and Ruan, Haifeng and Fan, Zhiyu and Roychoudhury, Abhik},
  booktitle={Proceedings of the 33rd ACM SIGSOFT International Symposium on Software Testing and Analysis},
  pages={1592--1604},
  year={2024}
}

@article{le2019automated,
  title={Automated program repair},
  author={Le Goues, Claire and Pradel, Michael and Roychoudhury, Abhik},
  journal={Communications of the ACM},
  volume={62},
  number={12},
  pages={56--65},
  year={2019},
  publisher={ACM New York, NY, USA}
}

@article{zhang2023survey,
  title={A survey of learning-based automated program repair},
  author={Zhang, Quanjun and Fang, Chunrong and Ma, Yuxiang and Sun, Weisong and Chen, Zhenyu},
  journal={ACM Transactions on Software Engineering and Methodology},
  volume={33},
  number={2},
  pages={1--69},
  year={2023},
  publisher={ACM New York, NY}
}

@article{huang2024evolving,
  title={Evolving paradigms in automated program repair: Taxonomy, challenges, and opportunities},
  author={Huang, Kai and Xu, Zhengzi and Yang, Su and Sun, Hongyu and Li, Xuejun and Yan, Zheng and Zhang, Yuqing},
  journal={ACM Computing Surveys},
  volume={57},
  number={2},
  pages={1--43},
  year={2024},
  publisher={ACM New York, NY}
}

@inproceedings{xia2023automated,
  title={Automated program repair in the era of large pre-trained language models},
  author={Xia, Chunqiu Steven and Wei, Yuxiang and Zhang, Lingming},
  booktitle={2023 IEEE/ACM 45th International Conference on Software Engineering (ICSE)},
  pages={1482--1494},
  year={2023},
  organization={IEEE}
}

@misc{darpa_aicyberchallenge_2024,
  author       = {{Defense Advanced Research Projects Agency (DARPA)}},
  title        = {DARPA AI Cyber Challenge},
  year         = {2024},
  howpublished = {\url{https://aicyberchallenge.com/}},
  note         = {Accessed: 2024-12-08}
}

@article{xia2025demystifying,
  title={Demystifying llm-based software engineering agents},
  author={Xia, Chunqiu Steven and Deng, Yinlin and Dunn, Soren and Zhang, Lingming},
  journal={Proceedings of the ACM on Software Engineering},
  volume={2},
  number={FSE},
  pages={801--824},
  year={2025},
  publisher={ACM New York, NY, USA}
}

@article{patil2024gorilla,
  title={Gorilla: Large language model connected with massive apis},
  author={Patil, Shishir G and Zhang, Tianjun and Wang, Xin and Gonzalez, Joseph E},
  journal={Advances in Neural Information Processing Systems},
  volume={37},
  pages={126544--126565},
  year={2024}
}

@article{schick2023toolformer,
  title={Toolformer: Language models can teach themselves to use tools},
  author={Schick, Timo and Dwivedi-Yu, Jane and Dess{\`\i}, Roberto and Raileanu, Roberta and Lomeli, Maria and Hambro, Eric and Zettlemoyer, Luke and Cancedda, Nicola and Scialom, Thomas},
  journal={Advances in Neural Information Processing Systems},
  volume={36},
  pages={68539--68551},
  year={2023}
}

@article{qin2023toolllm,
  title={Toolllm: Facilitating large language models to master 16000+ real-world apis},
  author={Qin, Yujia and Liang, Shihao and Ye, Yining and Zhu, Kunlun and Yan, Lan and Lu, Yaxi and Lin, Yankai and Cong, Xin and Tang, Xiangru and Qian, Bill and others},
  journal={arXiv preprint arXiv:2307.16789},
  year={2023}
}

@article{chen2025red,
  title={Red Teaming Program Repair Agents: When Correct Patches can Hide Vulnerabilities},
  author={Chen, Simin and He, Yixin and Jana, Suman and Ray, Baishakhi},
  journal={arXiv preprint arXiv:2509.25894},
  year={2025}
}

@article{lee2024unified,
  title={A unified debugging approach via llm-based multi-agent synergy},
  author={Lee, Cheryl and Xia, Chunqiu Steven and Yang, Longji and Huang, Jen-tse and Zhu, Zhouruixin and Zhang, Lingming and Lyu, Michael R},
  journal={arXiv preprint arXiv:2404.17153},
  year={2024}
}

@inproceedings{zhang2024poster,
  title={Poster: Repairing Bugs with the Introduction of New Variables: A Multi-Agent Large Language Model},
  author={Zhang, Elisa and Sun, Shiyu and Xing, Yunlong and Sun, Kun},
  booktitle={Proceedings of the 2024 on ACM SIGSAC Conference on Computer and Communications Security},
  pages={4961--4963},
  year={2024}
}

@misc{anthropic2024mcp,
  author       = {Anthropic},
  title        = {{Model Context Protocol (MCP) -- Open standard for integrating AI tools}},
  howpublished = {\url{https://ModelContextProtocol.io}},
  year         = {2024},
  note         = {Accessed: 2025-12-08}
}

@misc{anthropic_claudecode_2025,
  author = {Anthropic},
  title  = {{Claude Code} [AI coding assistant]},
  year   = {2025},
  howpublished = {\url{https://claude.ai/}},
  note   = {Accessed: YYYY-MM-DD}
}

@misc{github_copilot_2025,
  author = {GitHub},
  title  = {{GitHub Copilot} [AI pair programmer]},
  year   = {2025},
  howpublished = {\url{https://github.com/features/copilot}},
  note   = {Accessed: YYYY-MM-DD}
}

@article{chen2021codex,
  title={Evaluating Large Language Models Trained on Code},
  author={Chen, Mark and Tworek, Jerry and Jun, Heewoo and Yuan, Qiming and ...},
  journal={arXiv preprint arXiv:2107.03374},
  year={2021}
}

@article{afc-crs-all-you-need-is-a-fuzzing-brain,
  title={All You Need Is A Fuzzing Brain: An LLM-Powered System for Automated Vulnerability Detection and Patching},
  author={Sheng, Ze and Xu, Qingxiao and Huang, Jianwei and Woodcock, Matthew and Huang, Heqing and Donaldson, Alastair F and Gu, Guofei and Huang, Jeff},
  journal={arXiv preprint arXiv:2509.07225},
  year={2025}
}

@misc{aixcc-afc-archive,
  author       = {{theori-io}},
  title        = {{AIXCC AFC Archive}},
  howpublished = {\url{https://github.com/theori-io/aixcc-afc-archive}},
  year         = {2025},
  note         = {Accessed: 2025-12-18}
}

@misc{afc-buttercup,
  author       = {{trailofbits}},
  title        = {{AFC Buttercup}},
  howpublished = {\url{https://github.com/trailofbits/afc-buttercup}},
  year         = {2025},
  note         = {Accessed: 2025-12-18}
}

@article{kim2025atlantis,
  title={ATLANTIS: AI-driven Threat Localization, Analysis, and Triage Intelligence System},
  author={Kim, Taesoo and Han, HyungSeok and Park, Soyeon and Jeong, Dae R and Kim, Dohyeok and Kim, Dongkwan and Kim, Eunsoo and Kim, Jiho and Wang, Joshua and Kim, Kangsu and others},
  journal={arXiv preprint arXiv:2509.14589},
  year={2025}
}

@inproceedings{xiong2017precise,
  title={Precise condition synthesis for program repair},
  author={Xiong, Yingfei and Wang, Jie and Yan, Runfa and Zhang, Jiachen and Han, Shi and Huang, Gang and Zhang, Lu},
  booktitle={2017 IEEE/ACM 39th International Conference on Software Engineering (ICSE)},
  pages={416--426},
  year={2017},
  organization={IEEE}
}

@inproceedings{jiang2018shaping,
  title={Shaping program repair space with existing patches and similar code},
  author={Jiang, Jiajun and Xiong, Yingfei and Zhang, Hongyu and Gao, Qing and Chen, Xiangqun},
  booktitle={Proceedings of the 27th ACM SIGSOFT international symposium on software testing and analysis},
  pages={298--309},
  year={2018}
}

@misc{openai_aardvark2025,
  title        = {Introducing Aardvark: OpenAI's agentic security researcher},
  author       = {{OpenAI}},
  year         = {2025},
  month        = {Oct},
  day          = {30},
  howpublished = {\url{https://openai.com/index/introducing-aardvark/}},
  note         = {Accessed: 2026-01-12}
}

@misc{deepmind_codemender2025,
  author       = {Popa, Raluca Ada and Flynn, Fionn},
  title        = {Introducing CodeMender: An AI Agent for Code Security},
  year         = {2025},
  month        = {Oct},
  day          = {06},
  howpublished = {\url{https://deepmind.google/blog/introducing-codemender-an-ai-agent-for-code-security/}},
  note         = {DeepMind Blog; Accessed: 2026-01-12}
}

@ARTICLE{6035728,
  author={Le Goues, Claire and Nguyen, ThanhVu and Forrest, Stephanie and Weimer, Westley},
  journal={IEEE Transactions on Software Engineering}, 
  title={GenProg: A Generic Method for Automatic Software Repair}, 
  year={2012},
  volume={38},
  number={1},
  pages={54-72},
  doi={10.1109/TSE.2011.104}}

@inproceedings{10.1145/2837614.2837617,
author = {Long, Fan and Rinard, Martin},
title = {Automatic patch generation by learning correct code},
year = {2016},
isbn = {9781450335492},
publisher = {Association for Computing Machinery},
address = {New York, NY, USA},
url = {https://doi.org/10.1145/2837614.2837617},
doi = {10.1145/2837614.2837617},
booktitle = {Proceedings of the 43rd Annual ACM SIGPLAN-SIGACT Symposium on Principles of Programming Languages},
pages = {298–312},
numpages = {15},
location = {St. Petersburg, FL, USA},
series = {POPL '16}
}

@inproceedings{10.1145/2884781.2884807,
author = {Mechtaev, Sergey and Yi, Jooyong and Roychoudhury, Abhik},
title = {Angelix: scalable multiline program patch synthesis via symbolic analysis},
year = {2016},
isbn = {9781450339001},
publisher = {Association for Computing Machinery},
address = {New York, NY, USA},
url = {https://doi.org/10.1145/2884781.2884807},
doi = {10.1145/2884781.2884807},
booktitle = {Proceedings of the 38th International Conference on Software Engineering},
pages = {691–701},
numpages = {11},
location = {Austin, Texas},
series = {ICSE '16}
}

@INPROCEEDINGS{6606623,
  author={Nguyen, Hoang Duong Thien and Qi, Dawei and Roychoudhury, Abhik and Chandra, Satish},
  booktitle={2013 35th International Conference on Software Engineering (ICSE)}, 
  title={SemFix: Program repair via semantic analysis}, 
  year={2013},
  volume={},
  number={},
  pages={772-781},
  doi={10.1109/ICSE.2013.6606623}}

@article{yang2024swe,
  title={Swe-agent: Agent-computer interfaces enable automated software engineering},
  author={Yang, John and Jimenez, Carlos E and Wettig, Alexander and Lieret, Kilian and Yao, Shunyu and Narasimhan, Karthik and Press, Ofir},
  journal={Advances in Neural Information Processing Systems},
  volume={37},
  pages={50528--50652},
  year={2024}
}

@inproceedings{stephens2016driller,
  title={Driller: Augmenting fuzzing through selective symbolic execution.},
  author={Stephens, Nick and Grosen, John and Salls, Christopher and Dutcher, Andrew and Wang, Ruoyu and Corbetta, Jacopo and Shoshitaishvili, Yan and Kruegel, Christopher and Vigna, Giovanni},
  booktitle={NDSS},
  volume={16},
  number={2016},
  pages={1--16},
  year={2016}
}

@INPROCEEDINGS{6234425,
  author={Cha, Sang Kil and Avgerinos, Thanassis and Rebert, Alexandre and Brumley, David},
  booktitle={2012 IEEE Symposium on Security and Privacy}, 
  title={Unleashing Mayhem on Binary Code}, 
  year={2012},
  volume={},
  number={},
  pages={380-394},
  doi={10.1109/SP.2012.31}}

@misc{darpa_cgc,
  title        = {CGC: Cyber Grand Challenge},
  author       = {{Defense Advanced Research Projects Agency (DARPA)}},
  howpublished = {\url{https://www.darpa.mil/research/programs/cyber-grand-challenge}},
  year         = {2025},
  note         = {Accessed: 2026-01-12}
}

@inproceedings{zhou2024large,
  title={Large language model for vulnerability detection: Emerging results and future directions},
  author={Zhou, Xin and Zhang, Ting and Lo, David},
  booktitle={Proceedings of the 2024 ACM/IEEE 44th International Conference on Software Engineering: New Ideas and Emerging Results},
  pages={47--51},
  year={2024}
}

@inproceedings{guo2024outside,
  title={Outside the comfort zone: Analysing llm capabilities in software vulnerability detection},
  author={Guo, Yuejun and Patsakis, Constantinos and Hu, Qiang and Tang, Qiang and Casino, Fran},
  booktitle={European symposium on research in computer security},
  pages={271--289},
  year={2024},
  organization={Springer}
}

@article{sheng2025llms,
  title={Llms in software security: A survey of vulnerability detection techniques and insights},
  author={Sheng, Ze and Chen, Zhicheng and Gu, Shuning and Huang, Heqing and Gu, Guofei and Huang, Jeff},
  journal={ACM Computing Surveys},
  volume={58},
  number={5},
  pages={1--35},
  year={2025},
  publisher={ACM New York, NY}
}
